\documentclass[hyper]{JHEP3}
\usepackage{epsfig}
\usepackage{amsmath}
\usepackage{cite}

\newcommand{\beq}{\begin{equation}}
\newcommand{\eeq}{\end{equation}}
\newcommand{\bea}{\begin{eqnarray}}
\newcommand{\eea}{\end{eqnarray}}
\newcommand{\met}{\not \!\! E_T}
\newcommand{\etam}{\eta_{\rm max}}
\newcommand{\thc}{\theta_c}
\newcommand{\alps}{\alpha_{\rm S}}

\title{Effects of QCD radiation on inclusive variables for determining the scale of new physics at
hadron colliders}

\author{Andreas Papaefstathiou$^a$ and Bryan Webber$^{a,b}$\\
        $^a$Cavendish Laboratory, J.J.\ Thomson Avenue, Cambridge, UK\\
        $^b$Theory Group, Physics Department, CERN, 1211 Geneva 23, Switzerland\\
        E-mail:\email{andreas@hep.phy.cam.ac.uk},\email {webber@hep.phy.cam.ac.uk}
        }

\received{\today}               
\accepted{\today}               

\preprint{Cavendish-HEP-09/02\\CERN-PH-TH-2009-029\\MCnet/09/05}

\abstract{
We examine the effects of QCD initial-state radiation on a class of quantities, designed
to probe the mass scale of new physics at hadron colliders, which involve longitudinal as well
as transverse final-state momenta.  In particular, we derive universal functions that relate the
invariant mass and energy distribution of the visible part of the final state to that of the underlying
hard subprocess.  Knowledge of this relationship may assist in checking hypotheses about new processes,
by providing additional information about their scales. We compare our results with those of Monte
Carlo studies and find good general agreement.
}


\begin{document} 

\section{Introduction}
\label{sec:intro}
Searching for new physics at hadron colliders is difficult, partly because of the complexity of the expected
 new signals (typically, multiple jets and/or leptons plus missing energy) but also because processes at
 high mass scales are accompanied by copious initial-state QCD radiation (ISR).  As this tends to be
emitted close to the incoming beam directions, and the longitudinal momentum of the hard
subprocess is anyway unknown, search
variables~\cite{Hinchliffe:1996iu,Paige:1997xb,Lester:1999tx,Bachacou:1999zb,Hinchliffe:1999zc,
Tovey:2000wk,Allanach:2000kt,Barr:2003rg,Nojiri:2003tu,Kawagoe:2004rz,Gjelsten:2004ki,
Gjelsten:2005aw,Miller:2005zp,Lester:2006yw,Gjelsten:2006tg,Matsumoto:2006ws,Cheng:2007xv,
Lester:2007fq,Cho:2007qv,Gripaios:2007is,Barr:2007hy,Cho:2007dh,Ross:2007rm,Nojiri:2007pq,
Huang:2008ae,Nojiri:2008hy,Tovey:2008ui,Nojiri:2008ir,Cheng:2008mg,Cho:2008cu,Serna:2008zk,
Bisset:2008hm,Barr:2008ba,Kersting:2008qn,Nojiri:2008vq,Alwall:2008va,Cho:2008tj,Cheng:2008hk,
Burns:2008va,Barr:2008hv,Konar:2008ei} have generally been constructed either from the transverse
components of observed final-state momenta or else by assuming that some subset of these momenta
can be unambiguously ascribed to the hard subprocess.  Variables that make use of all observed momenta
without hypothesizing any particular structure of the final state are termed {\it global inclusive} variables
in ref.~\cite{Konar:2008ei}.  Examples of transverse global quantities of this type are the observed
transverse energy $E_T$, the missing transverse energy $\met$, and their sum $H_T=E_T+\met$.
The distributions of such quantities can provide information on the energy scales of new processes
such as supersymmetric particle production~\cite{Hinchliffe:1996iu,Paige:1997xb,Tovey:2000wk}.

Although the longitudinal components of final-state momenta are strongly influenced by ISR, they do
contain information about the underlying hard subprocess.  Indeed, the amount of ISR emitted is
determined by the energy scale of the subprocess.  It is therefore of interest to quantify the effects
of ISR on global observables that involve longitudinal momentum components.  The aim of the
present paper is to take the first steps in that direction.

In ref.~\cite{Konar:2008ei} various global search variables were investigated, including those that make
use of longitudinal as well as transverse momentum components.  The quantities studied included
the total energy $E$ visible in the detector and the visible invariant mass $M$,
\beq\label{eq:Mdef}
M = \sqrt{E^2-P_z^2-\met^2}
\eeq
where $P_z$ is the visible longitudinal momentum.  In addition a new variable was
introduced, defined as
\beq
\hat{s}^{1/2}_{\rm min}(M_{\rm inv}) \equiv \sqrt{M^2+\met^2}+\sqrt{M_{\rm inv}^2+\met^2}\ ,
\label{eq:smin_def}
\eeq
where the parameter $M_{\rm inv}$ is a variable estimating the sum of masses of all
invisible particles in the event:
\beq
M_{\rm inv} \equiv \sum_{i=1}^{n_{\rm inv}} m_i\ .
\label{eq:minv}
\eeq
It was argued that the peak in the distribution of $\hat{s}^{1/2}_{\rm min}$ is a good indicator
of the mass scale of new physics processes involving heavy particle production.

 It will also be useful to define the rapidity of the visible system,
\beq\label{eq:Ydef}
Y \equiv \frac 12 \ln\left(\frac{E+P_z}{E-P_z}\right)\ .
\eeq

The present paper examines the effects of ISR on global inclusive search variables, first in an
approximate fixed-order treatment taking into account collinear-enhanced terms, and then
in an all-orders resummation of such terms.  We quantify the way the distributions of
quantities that involve longitudinal momenta depend on the scale of the underlying hard
subprocess and on the properties of the detector, in particular the maximum visible
pseudorapidity $\etam$.   With the insight thus gained, it may be possible to correct for
this dependence and thereby extract information on the hard subprocess from such quantities.

\section{Fixed-order analysis}
The Monte Carlo results presented in ref.~\cite{Konar:2008ei} show that the second term on
the right-hand side of eq.~(\ref{eq:smin_def}) is not strongly affected by ISR. The first term is
intended to add extra longitudinal information about the hard subprocess, allowing a more reliable
determination of its mass scale. The extra longitudinal information enters through the visible mass $M$.
We therefore concentrate mainly on this quantity.

\subsection{Born approximation}
Since the effect of invisible final-state particles is not the central issue here, let us
suppose first that all the final-state particles from the hard subprocess are
detected.\footnote{We comment in sect.~5 on the treatment of invisible particles.}
Then in Born approximation, assuming that no beam remnants are detected, $M$ yields a perfect
estimate of the centre-of-mass energy of the hard subprocess.  For incoming partons with
momentum fractions $x_{1,2}$,
\beq\label{eq:EPBorn}
E=\frac 12 \sqrt{S} (x_1+x_2)\;,\;\;\;
P_z=\frac 12 \sqrt{S} (x_1-x_2)\;,
\eeq
where $\sqrt S$ is the hadron-hadron centre-of-mass energy, so that
\beq\label{eq:QYBorn}
M^2 = x_1x_2S\;,\;\;\;
Y=\frac 12 \ln\frac{x_1}{x_2}\ .
\eeq
The differential cross section for parton flavours $a,b$ is thus
\beq\label{eq:sigMY}
\frac{d\sigma_{ab}}{dM^2 dY} = \int dx_1\,dx_2\,f_a(x_1)f_b(x_2)\delta(M^2-x_1x_2S)
\delta\left(Y-\frac 12 \ln\frac{x_1}{x_2}\right)\hat\sigma_{ab}(x_1x_2S)
\eeq
where $f_{a,b}$ are the relevant parton distribution functions for the incoming hadrons and
$\hat\sigma_{ab}$ is the hard subprocess cross section. Hence at Born level we find
\beq
S\frac{d\sigma_{ab}}{dM^2 dY} =
f_a\left(\frac{M}{\sqrt S}e^{Y}\right)f_b\left(\frac{M}{\sqrt S}e^{-Y}\right)\hat\sigma_{ab}(M^2)\ .
\eeq

The parton distributions are normally given as $F_i(x)=xf_i(x)$, in terms of which we have
\beq\label{eq:dsigabBorn}
M^2\frac{d\sigma_{ab}}{dM^2 dY} =
F_a\left(\frac{M}{\sqrt S}e^{Y}\right)F_b\left(\frac{M}{\sqrt S}e^{-Y}\right)\hat\sigma_{ab}(M^2)\ .
\eeq
If the partonic cross section $\hat\sigma_{ab}$ has a threshold or peak, indicating that
the $ab$ subprocess has a characteristic scale $Q$, then this is also manifest
in the Born cross section (\ref{eq:dsigabBorn}) at $M\sim Q$, provided the relevant
parton distributions are large enough for that subprocess to contribute significantly. 

From eqs.~(\ref{eq:EPBorn}) and (\ref{eq:QYBorn}) we have $E=M\cosh Y$ and therefore the
visible energy distribution is given in Born approximation by
\beq\label{eq:dsigEBorn}
E^2\frac{d\sigma_{ab}}{dE^2 dY} =\left.M^2\frac{d\sigma_{ab}}{dM^2 dY}
\right|_{M=E\,\mbox{\scriptsize sech}\,Y}\ .
\eeq

\subsection{Quasi-collinear NLO correction}
To examine the sensitivity of the above results to ISR, let us first compute the NLO contribution due to
quasi-collinear gluon emission and the associated virtual corrections.  Consider first the emission of
a gluon from parton $a$.  If the emission angle $\theta$ is large enough, say $\theta>\thc$, the
gluon enters the detector and contributes to $M$. In the small-angle approximation we then have
\beq\label{eq:EPnlo}
E=\frac 12 \sqrt{S} (x_1/z+x_2)\;,\;\;\;
P_z=\frac 12 \sqrt{S} (x_1/z-x_2)\;,
\eeq
where $x_1/z$ is the momentum fraction of parton $a$ before the emission, so that
\beq\label{eq:MYnlo}
M^2= x_1x_2S/z\;,\;\;\;
Y=\frac 12 \ln\frac{x_1}{zx_2}\ .
\eeq
The correction associated with a detected emission from parton $a$ is then
\beq
\frac{\alps}\pi\int_{\thc}\frac{d\theta}\theta \frac{dz}z dx_1\,dx_2
\tilde P_a(z)f_a(x_1/z)f_b(x_2)\delta(M^2-x_1x_2S/z)
\delta\left(Y-\frac 12 \ln\frac{x_1}{zx_2}\right)\hat\sigma_{ab}(x_1x_2S)
\eeq
 where $\tilde P_a(z)$ is the unregularized $a\to ag$ splitting function.

On the other hand if the gluon misses the detector ($\theta<\thc$),
$E$ and $P_z$ are still given by (\ref{eq:EPBorn}), so the contribution is
\beq
\frac{\alps}\pi\int_0^{\thc}\frac{d\theta}\theta \frac{dz}z dx_1\,dx_2
\tilde P_a(z)f_a(x_1/z)f_b(x_2)\delta(M^2-x_1x_2S)
\delta\left(Y-\frac 12 \ln\frac{x_1}{x_2}\right)\hat\sigma_{ab}(x_1x_2S)\,.
\eeq
Finally the associated virtual correction is the term that regularizes the splitting function,
which in this approximation is
\beq
-\frac{\alps}\pi\int\frac{d\theta}\theta\,dz\,dx_1\,dx_2
\tilde P_a(z)f_a(x_1)f_b(x_2)\delta(M^2-x_1x_2S)
\delta\left(Y-\frac 12 \ln\frac{x_1}{x_2}\right)\hat\sigma_{ab}(x_1x_2S)\,.
\eeq
Adding everything together gives a correction
\bea
\delta\left(\frac{d\sigma_{ab}}{dM^2 dY}\right) &=&\frac{\alps}\pi
\int\frac{d\theta}\theta\,dz\,dx_1\,dx_2\,\tilde P_a(z)f_b(x_2)
\hat\sigma_{ab}(x_1x_2S)\nonumber\\
&&\Bigl[\frac 1z f_a(x_1/z)\delta\left(Y-\frac 12 \ln\frac{x_1}{zx_2}\right)
\delta(M^2-x_1x_2S/z)\Theta(\theta-\thc)\\
&&+\Bigl\{\frac 1z f_a(x_1/z)\Theta(\thc-\theta)
-f_a(x_1)\Bigr\}\delta\left(Y-\frac 12 \ln\frac{x_1}{x_2}\right)
\delta(M^2-x_1x_2S)\Bigr]\ .\nonumber
\eea

Setting aside for the moment the possibility of splittings other than $a\to ag$,
the DGLAP evolution equation for $f_a(x_1)$ is
\beq\label{eq:evol}
q\frac{\partial}{\partial q}f_a(x_1) =\frac{\alps}\pi\int dz\,\tilde P_a(z)
\left[\frac 1z f_a(x_1/z) - f_a(x_1)\right]
\eeq
where $q$ represents the scale at which the parton distribution is measured.
Hence the correction may be written as
\bea
\delta\left(\frac{d\sigma_{ab}}{dM^2dY}\right) &=&
\int\frac{d\theta}\theta\,dx_1\,dx_2\,f_b(x_2)\hat\sigma_{ab}(x_1x_2S)
\Biggl[q\frac{\partial f_a}{\partial q}\delta\left(Y-\frac 12 \ln\frac{x_1}{x_2}
\right)\delta(M^2-x_1x_2S)\nonumber\\
&&+\frac{\alps}\pi\int\frac{dz}z \tilde P_a(z)\,f_a(x_1/z)\Biggl\{\delta\left(Y-\frac 12 \ln\frac{x_1}{zx_2}\right)\delta(M^2-x_1x_2S/z)\nonumber\\
&& -\delta\left(Y-\frac 12 \ln\frac{x_1}{x_2}\right)\delta(M^2-x_1x_2S)
\Biggr\}\Theta(\theta-\thc)\Biggr]\ .
\eea
Since $d\theta/\theta = dq/q$, the first term represents a change of scale in the Born term.
It replaces the reference scale in $f_a$ by the scale $Q$ of the hard subprocess. 
 The remaining terms give a correction
\bea
\delta\left(\frac{d\sigma_{ab}}{dM^2dY}\right) &=&
\frac{\alps}{\pi S}\int_{\thc}\frac{d\theta}\theta\int\frac{dz}z \tilde P_a(z)
f_b\left(\frac{M}{\sqrt S}e^{-Y}\right)\times\nonumber\\
&&\left[f_a\left(\frac{M}{\sqrt S}e^{Y}\right)z\hat\sigma_{ab}(zM^2)
-f_a\left(\frac{M}{z\sqrt S}e^{Y}\right)\hat\sigma_{ab}(M^2)\right]\ .
\eea
In leading-log approximation the $\theta$ integration just gives a factor of $-\ln\thc$.
In the same approximation, we may set $-\ln\thc = \etam$, the maximum
pseudorapidity seen by the detector.  Note that this is a different quantity from $Y$,
the true rapidity of the visible system, given by eq.~(\ref{eq:Ydef}).
The correction associated with parton $b$ gives the same expression with
$a\leftrightarrow b$ and $Y\to -Y$.  Thus, defining
\beq
\bar x_1 = \frac{M}{\sqrt S}e^{Y}\;,\;\;\;
\bar x_2 = \frac{M}{\sqrt S}e^{-Y}\;,
\eeq
we have
\bea
S\frac{d\sigma_{ab}}{dM^2 dY} &=&
f_a(\bar x_1,Q)f_b(\bar x_2,Q)\hat\sigma_{ab}(M^2)\nonumber\\
&+&\etam\frac{\alps}\pi\int\frac{dz}z
\Bigl[z\{\tilde P_a(z)+\tilde P_b(z)\}f_a(\bar x_1,Q)f_b(\bar x_2,Q)
\hat\sigma_{ab}(zM^2)\nonumber\\
&-&\{\tilde P_a(z)f_a(\bar x_1/z,Q)f_b(\bar x_2,Q)+
\tilde P_b(z)f_a(\bar x_1,Q)f_b(\bar x_2/z,Q)\}\hat\sigma_{ab}(M^2)\Bigr].
\eea
Expressing this in terms of $F_i(x)=xf_i(x)$, as in eq.~(\ref{eq:dsigabBorn}),
\bea\label{eq:M2sig}
M^2\frac{d\sigma_{ab}}{dM^2 dY} &=&
F_a(\bar x_1,Q)F_b(\bar x_2,Q)\hat\sigma_{ab}(M^2)\nonumber\\
&+&\etam\frac{\alps}\pi\int dz
\Bigl[\{\tilde P_a(z)+\tilde P_b(z)\}F_a(\bar x_1,Q)F_b(\bar x_2,Q)\hat\sigma_{ab}(zM^2)\\
&-&\{\tilde P_a(z)F_a(\bar x_1/z,Q)F_b(\bar x_2,Q)+\tilde P_b(z)
F_a(\bar x_1,Q)F_b(\bar x_2/z,Q)\}\hat\sigma_{ab}(M^2)\Bigr]\ .\nonumber
\eea

\begin{figure}
\begin{center}
\epsfig{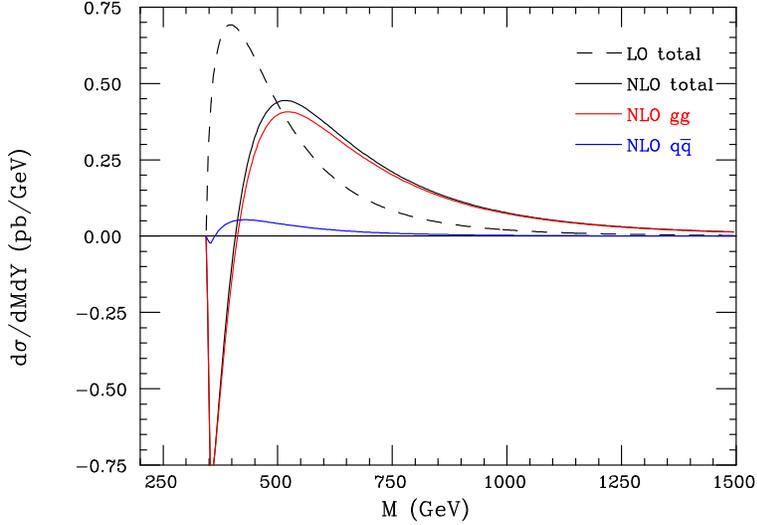}
\end{center}
\caption{Distribution of visible mass $M$ in $t\bar t$ production at LHC for $\etam=5$ and
$Y=0$: leading and approximate next-to-leading order.\label{fig:ttbar}}
\end{figure}

Results for $t\bar t$ production at the LHC ($pp$ at $\sqrt S=14$ TeV) with $\etam=5$ and $Y=0$
are shown in fig.~\ref{fig:ttbar}.  Leading-order MSTW parton distributions~\cite{Martin:2009iq}
were used.  For simplicity we have taken $Q=M$.
Recall that the simplifying assumption made here is that all $t\bar t$ decay products are detected,
so the $M$ distribution vanishes below $t\bar t$ threshold.  We see that there is a large negative
NLO correction near threshold, followed by a broad positive peak.

To understand these qualitative features, consider the case $a=b$, as in $gg\to t\bar t$,
and $Y=0$, so that $\bar x_{1,2} = M/\sqrt S\equiv \bar x$. Then the NLO correction
becomes simply
\beq
\delta\left(M^2\frac{d\sigma}{dM^2dY}\right) =
2\etam\frac{\alps}\pi F(\bar x) \int dz\tilde
P_a(z)\left[F(\bar x)\hat\sigma(zM^2)-F(\bar x/z)\hat\sigma(M^2)\right]
\eeq
The first term is positive-definite, contributes only above threshold,
and diverges at threshold.  It produces the broad positive peak.  The
second term is negative-definite, contributes around threshold, and
has a divergent coefficient. It provides the sharp negative peak.

From eqs.~(\ref{eq:EPnlo}) and (\ref{eq:MYnlo}), the relation $E=M\cosh Y$ still holds in this
approximation and therefore the visible energy distribution is again given by eq.~(\ref{eq:dsigEBorn}).

\section{Resummation}
By adding and subtracting the expression
\beq
\{\tilde P_a(z)+\tilde P_b(z)\}F_a(\bar x_1,Q)F_b(\bar x_2,Q)\hat\sigma_{ab}(M^2)
\eeq
in the integrand of eq.~(\ref{eq:M2sig}) and comparing with eq.~(\ref{eq:evol}), we see that the last line of 
that equation corresponds to a change of scale $Q\to Q_c=\thc Q$ in the parton distributions, leading to
\beq\label{eq:M2sigFFS}
M^2\frac{d\sigma_{ab}}{dM^2 dY} = F_a(\bar x_1,Q_c)F_b(\bar x_2,Q_c) \Sigma_{ab}(M^2)
\eeq
where to first order
\bea\label{eq:SigOa}
\Sigma_{ab}(M^2) &=& \hat\sigma_{ab}(M^2)\nonumber\\
&+&\etam\frac{\alps}\pi\int dz\{\tilde P_a(z)+\tilde P_b(z)\}\{\hat\sigma_{ab}(zM^2)-\hat\sigma_{ab}(M^2)\}\;.
\eea
The interpretation of this result is simple:  undetected ISR at angles less than $\thc$, corresponding to scales less
than $\thc Q$, is absorbed into the structure of the incoming hadrons.  To resum the effects of gluons at angles
greater than $\thc$, consider first the real emission of $n$ such gluons from parton $a$.  In the quasi-collinear
approximation these form an angular-orders sequence, giving rise to a contribution to $\Sigma_{ab}$ of
\bea\label{eq:nreal}
&&\left(\frac{\alps}\pi\right)^n\int_{\thc}\frac{d\theta_1}{\theta_1}\int_{\theta_1}\frac{d\theta_2}{\theta_2}
\ldots \int_{\theta_{n-1}}\frac{d\theta_n}{\theta_n}\int_0^1 dz_1\ldots dz_n \tilde P_a(z_1)\ldots \tilde P_a(z_n)\hat\sigma_{ab}(z_1\ldots z_n M^2)\nonumber\\
&&=\frac 1{n!}\left(\etam\frac{\alps}\pi\right)^n
\int_0^1 dz_1\ldots dz_n \tilde P_a(z_1)\ldots \tilde P_a(z_n)\hat\sigma_{ab}(z_1\ldots z_n M^2)\;,
\eea
where again we have made the identification $-\ln\thc = \etam$.
The multiple convolution of the momentum fractions $z_i$ can be transformed into a product by taking moments.
Defining
\beq\label{eq:momdef}
\int_0^\infty dM^2 \left(M^2\right)^{-N} \hat\sigma_{ab}(M^2) \equiv \hat\sigma^{ab}_N
\eeq
we have
\beq
\left(\frac{\alps}\pi\right)^n
\int_0^\infty dM^2 \left(M^2\right)^{-N} \int dz_1\ldots dz_n \tilde P_a(z_1)\ldots  \tilde P_a(z_n)
\hat\sigma(z_1\ldots z_n M^2) = \left(\tilde\gamma^a_N\right)^n\hat\sigma^{ab}_N
\eeq
where
\beq
\tilde\gamma^a_N = \frac{\alps}\pi\int_0^1 dz\,z^{N-1}\tilde P_a(z)\ .
\eeq
Therefore defining correspondingly
\beq\label{eq:SigNdef}
\int_0^\infty dM^2 \left(M^2\right)^{-N} \Sigma_{ab}(M^2) \equiv \Sigma^{ab}_N\;,
\eeq
the contribution (\ref{eq:nreal}) to this quantity will be
\beq
\frac 1{n!}\left(\etam\tilde\gamma^a_N\right)^n \hat\sigma^{ab}_N
\eeq
which summed over $n$ gives
\beq
\exp\left(\etam\tilde\gamma^a_N\right)\hat\sigma^{ab}_N\ .
\eeq

The corresponding virtual contributions give a Sudakov-like form factor
\beq
\exp\left(-\frac{\alps}\pi\int_{\thc}\frac{d\theta}\theta \int_0^1 dz\, \tilde P_a(z)\right)
\eeq
and therefore the total contribution from parton $a$ is
\beq
\exp\left(\etam\gamma^a_N\right)\hat\sigma^{ab}_N
\eeq
where $\gamma^a_N$ is the anomalous dimension
\beq
\gamma^a_N = \frac{\alps}\pi\int_0^1 dz\left(z^{N-1}-1\right)\tilde P_a(z)
=\frac{\alps}\pi\int_0^1 dz\,z^{N-1} P_a(z)\;,
\eeq
$P_a(z)$ being the regularized $a\to ag$ splitting function.
Parton $b$ gives a similar factor with $\gamma^b_N$ in place of $\gamma^a_N$, so the
result for the quantity  (\ref{eq:SigNdef}) is simply
\beq\label{eq:SigN}
\Sigma^{ab}_N  = e^{\etam(\gamma^a_N+\gamma^b_N)}\hat\sigma^{ab}_N\;.
\eeq
We can see as follows that this result is qualitatively correct.  The anomalous dimensions are positive
for small $N$ and negative for large $N$.  Thus, for $\thc\ll 1$, $\Sigma^{ab}_N$ is enhanced relative
to $\hat\sigma^{ab}_N$ at small $N$ and suppressed at large $N$.   Now from the moment definition
(\ref{eq:momdef}) small $N$ corresponds to large $M$ and vice versa.  Hence the distribution of $M$
is suppressed at small $M$ and enhanced at large $M$ relative to the Born term, as observed in the
Monte Carlo~\cite{Konar:2008ei} and NLO results.

The emission of partons other than gluons is included by introducing the anomalous dimension
matrix $\Gamma_N$ with elements given by
\beq
(\Gamma_N)_{ba} = \frac{\alps}\pi\int_0^1 dz\,z^{N-1}P_{ba}(z)
\eeq
where $P_{ba}(z)$ is the regularized $a\to b$ splitting function.  Then
\beq\label{eq:SigabN}
\Sigma^{ab}_N  =  \hat\sigma^{a'b'}_N\left(e^{\etam\Gamma_N}\right)_{a'a}
\left(e^{\etam\Gamma_N}\right)_{b'b}\ .
\eeq
The corresponding generalization of the evolution equation (\ref{eq:evol}) is
\beq\label{eq:evolba}
q\frac{\partial}{\partial q}f_b(x) =\frac{\alps}\pi\int\frac{dz}z P_{ba}(z) f_a(x/z)
\eeq
Defining the moments of the parton distribution functions
\beq
f^a_N = \int_0^1 dx\,x^{N-1} f_a(x)
\eeq
we see that
\beq\label{eq:evolN}
q\frac{\partial}{\partial q}f^b_N= (\Gamma_N)_{ba} f^a_N
\eeq
with solution
\beq
f^b_N(q) = \left([q/q_0]^{\Gamma_N}\right)_{ba} f^a_N(q_0)\ .
\eeq
Hence
\beq\label{eq:fbN}
f^b_N(Q) = \left(e^{\etam\Gamma_N}\right)_{ba} f^a_N(Q_c)\;,
\eeq
where
\beq\label{eq:Qc}
Q_c=\thc Q= Qe^{-\etam}\;,
\eeq
showing that the evolution of the visible mass distribution
 is related to that of the parton distributions over the same range of scales.

Taking into account the running of the strong coupling $\alps(q)$ in the
evolution equation (\ref{eq:evolba}), eq.~(\ref{eq:fbN}) becomes
\beq\label{eq:fbNK}
f^b_N(Q) = K^{ba}_N f^a_N(Q_c)
\eeq
where
\beq\label{eq:Kba}
 K^{ba}_N = \left(\left[\frac{\alps(Q_c)}{\alps(Q)}\right]^{p\Delta_N}\right)_{ba}
\eeq
with $p = 6/(11C_A-2n_f)$ and
\beq
(\Delta_N)_{ba} = \frac\pi{\alps}(\Gamma_N)_{ba} = \int_0^1 dz\,z^{N-1}P_{ba}(z) \ .
\eeq
The running of $\alps$ will affect eq.~(\ref{eq:SigabN}) similarly, giving
\beq\label{eq:SigabNK}
\Sigma^{ab}_N  =  \hat\sigma^{a'b'}_N K^{a'a}_N K^{b'b}_N\ .
\eeq
This implies that
\beq\label{eq:SigabH}
\Sigma_{ab}(M^2) =  \int_0^1 dz\,\hat\sigma_{a'b'}(zM^2)\, H_{a'b',ab}(z)
\eeq
where
\beq
K^{a'a}_N K^{b'b}_N = \int_0^1 dz\,z^{N-1}H_{a'b',ab}(z)
\eeq
or, inverting the Mellin transformation,
\beq\label{eq:Hdef}
H_{a'b',ab}(z) =\frac 1{2\pi i}\int_C dN\,z^{-N} K^{a'a}_N K^{b'b}_N
\eeq
where the contour $C$ is to the right of all singularities of the integrand.

Alternatively, eq.~(\ref{eq:SigabH}) can be expressed as a double convolution,
\beq\label{eq:SigabKK}
\Sigma_{ab}(M^2) =  \int_0^1 dz_1\,dz_2\,\hat\sigma_{a'b'}(z_1z_2M^2)\, K_{a'a}(z_1)\, K_{b'b}(z_2)
\eeq
where
\beq
K^{b'b}_N = \int_0^1 dz\,z^{N-1}K_{b'b}(z)\;.
\eeq
It then follows from eq.~(\ref{eq:Kba}) that $K_{b'b}(z)$ obeys an evolution equation like that of the parton distributions:
\beq\label{eq:evolK}
Q\frac{\partial}{\partial Q}K_{b'b}(z) =\frac{\alps(Q)}\pi\int\frac{dz'}{z'} P_{b'a}(z') K_{ab}(z/z')\;.
\eeq

Putting everything together, the visible mass distribution is related to the hard subprocess
cross section (in the absence of invisible final-state particles) as follows:
\beq\label{eq:M2sigz1z2}
M^2\frac{d\sigma_{ab}}{dM^2 dY} = \int dz_1\,dz_2\,\hat\sigma_{a'b'}(z_1 z_2 M^2)\,K_{a'a}(z_1)
F_a(\bar x_1,Q_c)\,K_{b'b}(z_2)F_b(\bar x_2,Q_c)
\eeq
where the kernel functions $K_{a'a}(z)$ and $K_{b'b}(z)$ can be obtained by solving the evolution
equation (\ref{eq:evolK}) with the initial condition that $K_{ab}(z)=\delta_{ab}\delta(1-z)$ at $Q=Q_c$.

Since the partons sampled at scale $Q_c$ are always regarded as (anti-)collinear, the relation
$E=M\cosh Y$ still holds and therefore the visible energy distribution is given by
\beq\label{eq:dsigEresum}
E^2\frac{d\sigma_{ab}}{dE^2 dY} =\left.M^2\frac{d\sigma_{ab}}{dM^2 dY}
\right|_{M=E\,\mbox{\scriptsize sech}\,Y}\ .
\eeq
in leading-logarithmic approximation to all orders.

To verify that the integrated cross section is not affected by resummation, define
$x_{1,2}=z_{1,2}\bar x_{1,2}$ and write eq.~(\ref{eq:M2sigz1z2}) as
\beq\label{eq:M2sigx1x2}
M^2\frac{d\sigma_{ab}}{dM^2 dY} = \int dx_1\,dx_2\,\hat\sigma_{a'b'}(x_1x_2S)\,K_{a'a}(x_1/\bar x_1)
f_a(\bar x_1,Q_c)\,K_{b'b}(x_2/\bar x_2)f_b(\bar x_2,Q_c)\ .
\eeq
Now
\beq
\frac{dM^2}{M^2}dY = \frac{d\bar x_1}{\bar x_1}\frac{d\bar x_2}{\bar x_2}
\eeq
and
\beq
\sum_a\int\frac{d\bar x_1}{\bar x_1}K_{a'a}(x_1/\bar x_1)f_a(\bar x_1,Q_c) = f_{a'}(x_1,Q)\ .
\eeq
Hence
\beq\label{eq:intMY}
\sum_{ab}\int dM^2dY\frac{d\sigma_{ab}}{dM^2 dY}
= \sum_{a'b'}\int dx_1\,dx_2\,\hat\sigma_{a'b'}(x_1x_2S) f_{a'}(x_1,Q)f_{b'}(x_2,Q)
\eeq
in agreement with eq.~(\ref{eq:sigMY}).

\begin{figure}
\begin{center}
\epsfig{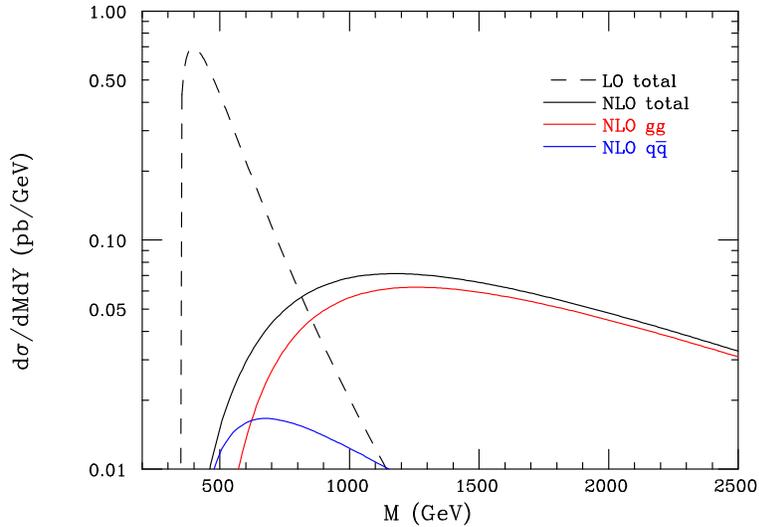}
\end{center}
\caption{Resummed distribution of visible mass $M$ in $t\bar t$ production at LHC for $\etam=5$
and $Y=0$.\label{fig:tresum}}
\end{figure}

\begin{figure}
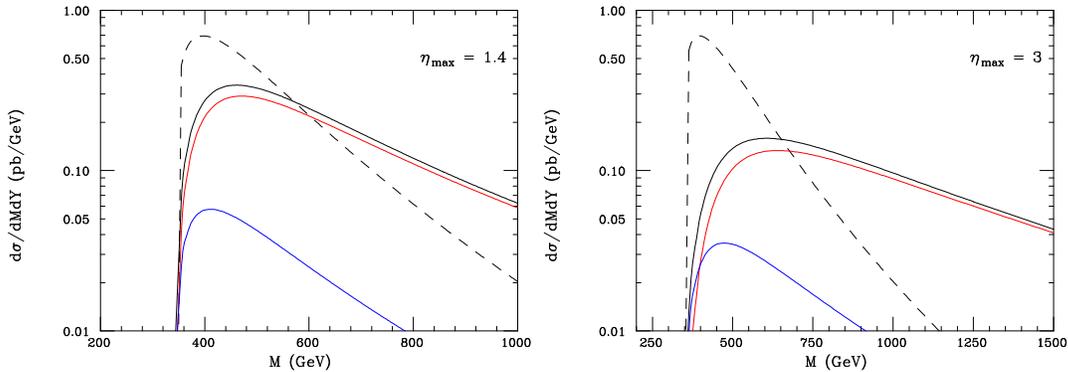

\begin{center}
\epsfig{file=mresumEta1p4.ps,width=7cm}
\epsfig{file=mresumEta3.ps,width=7cm}
\end{center}
\caption{Resummed distribution of visible mass $M$ in $t\bar t$ production at LHC for $Y=0$
and lower values of $\etam$: colour scheme as in figs.1 and 2.\label{fig:tresumEta}}
\end{figure}

Resummed results corresponding to fig.~\ref{fig:ttbar} are shown in fig.~\ref{fig:tresum}.  We see
that the peak of the distribution has moved to much higher mass, beyond 1 TeV.  This is due to
multiple emission of ISR partons in the evolution of the initial state from the detection scale $Q_c$
to the hard subprocess scale $Q$.  As the value of $\etam$ is reduced, the range of evolution becomes
smaller, less ISR is emitted, and the peak moves closer to the hard subprocess scale, as illustrated in
fig.~\ref{fig:tresumEta}.  However, recall that any loss of visible particles produced in the hard subprocess
(top decay products in this case) has been neglected here.  Such losses will cause the peak to fall
below the hard subprocess scale at low values of $\etam$.  In ref.~\cite{Konar:2008ei} it was found that
when $\etam=1.4$ the peak lies close to threshold for the hard subprocesses studied there, presumably
as a result of compensation between ISR and loss of visible decay products.

\begin{figure}
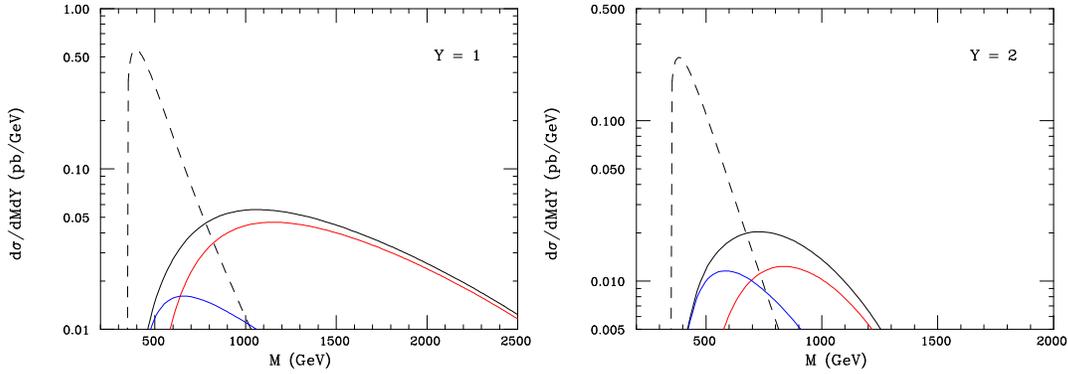

\begin{center}
\epsfig{file=mresumY1.ps,width=7cm}
\epsfig{file=mresumY2.ps,width=7cm}
\end{center}
\caption{Resummed distribution of visible mass $M$ in $t\bar t$ production at LHC for $\etam=5$:
results at non-zero visible rapidity $Y$.\label{fig:tresumY}}
\end{figure}

Results for higher values of the visible rapidity $Y$ are shown in fig.~\ref{fig:tresumY}.  The peak moves to
lower mass as $Y$ increases, as a consequence of the suppression of high masses by the rapid fall-off of
the parton distributions at high $x$.

\section{Monte Carlo comparisons}
\begin{figure}
\begin{center}
\epsfig{file=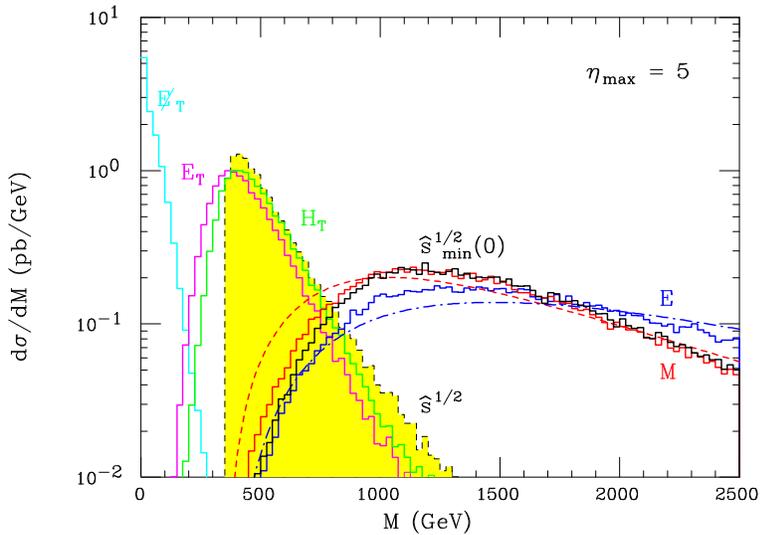,width=10cm}
\end{center}
\caption{Monte Carlo results for various inclusive observables in $t\bar t$ production
at LHC. The yellow filled histogram shows the true hard subprocess invariant mass.  The dashed
red and dot-dashed blue curves show the results of eqs.~({\protect\ref{eq:M2sigz1z2}})
and ({\protect\ref{eq:dsigEresum}}) , respectively, integrated with respect to $Y$.\label{fig:HerMC}}
\end{figure}

In this section we compare the predictions of the above analytical treatment with Monte Carlo results from
{\tt HERWIG}~\cite{Corcella:2000bw,Corcella:2002jc}.   Figure~\ref{fig:HerMC} shows the Monte Carlo results
for various global inclusive observables in  $t\bar t$ production at the LHC when $\etam=5$.  For the
detector simulation we used GETJET~\cite{GETJET} with calorimeter cell size
$(\Delta\eta,\Delta\phi)=(0.1,0.1)$.  To facilitate the study of ISR effects,
the simulated underlying event was turned off.

As expected, the distributions of variables that involve longitudinal momenta are broadened
and shifted relative to those of purely transverse quantities.  The distributions of the visible
mass $M$ and the new quantity (\ref{eq:smin_def}) are similar, while the visible energy $E$
has a broader distribution, as it includes the visible momentum.

Comparing the visible mass and energy distributions with the analytical calculations (the dashed red and
dot-dashed blue curves, respectively), we find good overall agreement, considering the
simplifications made in the latter, viz.\ the quasi-collinear approximation,  no loss of top decay products
and no hadronization.  In addition, turning off the underlying event in the Monte Carlo does not entirely
eliminate contributions from spectator parton fragments at high rapidity.

\begin{figure}
\begin{center}
\epsfig{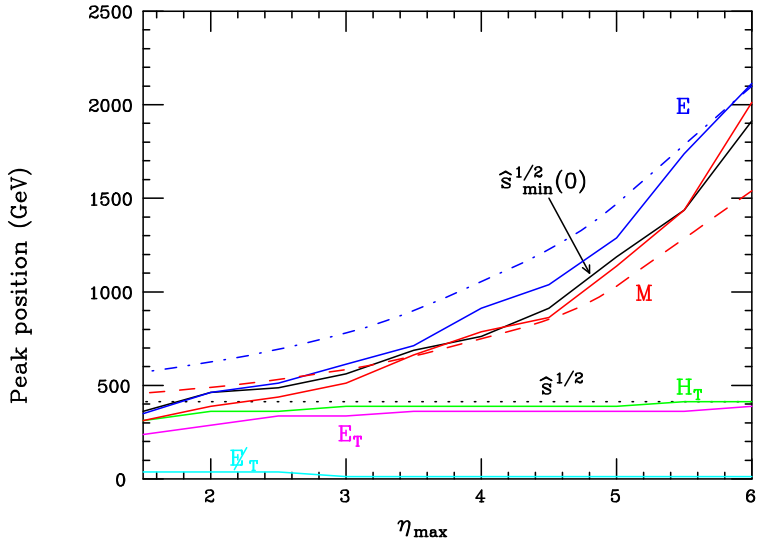}
\end{center}
\caption{Position of peaks in the distributions of various inclusive observables in $t\bar t$
production at LHC,  as a function of the maximum visible pseudorapidity $\etam$.
Solid curves: Monte Carlo results. Dashed red and dot-dashed blue curves: $M$ and $E$ peaks
from eqs.~({\protect\ref{eq:M2sigz1z2}}) and ({\protect\ref{eq:dsigEresum}}), respectively.
Dotted line: $t\bar t$ invariant mass peak.\label{fig:tresumpeak}}
\end{figure}

The motion of the peaks in these distributions as the  maximum visible pseudorapidity $\etam$ is
varied is shown in  fig.~\ref{fig:tresumpeak}.  The distributions of the transverse quantities remain
remarkably stable, while the peak positions of the other variables rise faster than linearly with
increasing $\etam$.  The analytical prediction for the peak position of the visible mass is in good
agreement with the Monte Carlo results at intermediate values of $\etam$ but somewhat higher at
low values and lower at high values.  This is consistent with the loss of top decay products
at low $\etam$ and the contribution of spectator fragments at high $\etam$ in the Monte Carlo.
The peak position of the visible energy distribution is somewhat overestimated at all but the
highest values of $\etam$.  However, as may be seen from fig.~\ref{fig:HerMC}, the distribution
of this quantity is quite flat near the peak, and therefore a small discrepancy between the analytical
and Monte Carlo results can give rise to a larger difference in the associated peak positions.

More precise Monte Carlo studies of global inclusive observables like those considered here could be
performed using methods that match parton showers to matrix elements for extra jet production; see for
example \cite{Mangano:2006rw,Alwall:2008qv,Hoeche:2009rj} and references therein.  We checked that
matching to NLO (up to one extra jet) using MC@NLO \cite{Frixione:2002ik,Frixione:2003ei} did not lead
to significantly different results.

\section{Conclusions}
We have derived a simple analytical formula, eq.~(\ref{eq:M2sigz1z2}), for the effects of QCD initial
state radiation on the invariant mass $M$ and rapidity $Y$ of the visible part of the final state
of a hadronic collision, as a function of the pseudorapidity coverage $\etam$ of an idealized detector.
Given the mass distribution of a hard subprocess involving incoming partons $a'$ and $b'$, a simple
convolution with the universal functions $H_{a'b',ab}$ in eq.~(\ref{eq:Hdef}) provides the quantities
$\Sigma_{ab}$ that, when multiplied by the relevant parton distributions, sampled at an
$\etam$-dependent scale, yield the visible mass and energy distributions.

The derivation involved a number of simplifying assumptions and approximations, but was in
satisfactory agreement with Monte Carlo results that should in principle be more realistic.
The difference between the visible mass $M$ and the new variable (\ref{eq:smin_def}) is small
when $M_{\rm inv}=0$, essentially because of the relative smallness of the missing transverse
energy $\met$, manifest in fig.~\ref{fig:HerMC}.

Many of the approximations made here could be improved if a more precise analysis is
needed.  The emission of invisible particles in the hard subprocess could be taken into
account by replacing the mass distribution of that process by a visible mass distribution, with
the invisible component already integrated out.  The universal functions $H_{a'b',ab}$ could
be computed to next-to-leading order, as they involve the same quantities that drive
the evolution of the parton distribution functions.  Together with an NLO calculation of
the hard subprocess mass distribution, this would provide a complete NLO description of
the visible mass and energy distributions.

Whether such refinements are worthwhile depends on the extent to which observables
involving longitudinal momenta are found useful in the exploration of physics beyond
the Standard Model.  The Monte Carlo results presented in the previous section
confirm that transverse quantities are much less sensitive to the effects of ISR.
However, at the very least it will be useful to check the consistency of hypotheses
about new subprocesses with the distributions discussed here, which do
contain independent information about their scales.

\section*{Acknowledgements}
We are grateful to Konstantin Matchev, Jennifer Smillie and members of the Cambridge Supersymmetry
Working Group for comments and discussion.  This work was supported in part by the UK Science and
Technology Facilities Council and the European Union Marie Curie Research Training Network
MCnet (contract MRTN-CT-2006-035606).



\begin{thebibliography}{999}

\bibitem{Hinchliffe:1996iu}
  I.~Hinchliffe, F.~E.~Paige, M.~D.~Shapiro, J.~Soderqvist and W.~Yao,
  ``Precision SUSY measurements at LHC,''
  Phys.\ Rev.\  D {\bf 55}, 5520 (1997)
  [arXiv:hep-ph/9610544].

\bibitem{Paige:1997xb}
  F.~E.~Paige,
  ``Supersymmetry signatures at the CERN LHC,''
  arXiv:hep-ph/9801254.

\bibitem{Lester:1999tx}
  C.~G.~Lester and D.~J.~Summers,
   ``Measuring masses of semi-invisibly decaying particles pair produced at
  hadron colliders,''
  Phys.\ Lett.\  B {\bf 463}, 99 (1999)
  [arXiv:hep-ph/9906349].

\bibitem{Bachacou:1999zb}
  H.~Bachacou, I.~Hinchliffe and F.~E.~Paige,
  ``Measurements of masses in SUGRA models at LHC,''
  Phys.\ Rev.\  D {\bf 62}, 015009 (2000)
  [arXiv:hep-ph/9907518].

\bibitem{Hinchliffe:1999zc}
  I.~Hinchliffe and F.~E.~Paige,
  ``Measurements in SUGRA models with large tan(beta) at LHC,''
  Phys.\ Rev.\  D {\bf 61}, 095011 (2000)
  [arXiv:hep-ph/9907519].

\bibitem{Tovey:2000wk}
  D.~R.~Tovey,
  ``Measuring the SUSY mass scale at the LHC,''
  Phys.\ Lett.\  B {\bf 498}, 1 (2001)
  [arXiv:hep-ph/0006276].

\bibitem{Allanach:2000kt}
  B.~C.~Allanach, C.~G.~Lester, M.~A.~Parker and B.~R.~Webber,
   ``Measuring sparticle masses in non-universal string inspired models at  the
  LHC,''
  JHEP {\bf 0009}, 004 (2000)
  [arXiv:hep-ph/0007009].

\bibitem{Barr:2003rg}
  A.~Barr, C.~Lester and P.~Stephens,
  ``m(T2): The truth behind the glamour,''
  J.\ Phys.\ G {\bf 29}, 2343 (2003)
  [arXiv:hep-ph/0304226].

\bibitem{Nojiri:2003tu}
  M.~M.~Nojiri, G.~Polesello and D.~R.~Tovey,
   ``Proposal for a new reconstruction technique for SUSY processes at the
  LHC,''
  arXiv:hep-ph/0312317.

\bibitem{Kawagoe:2004rz}
  K.~Kawagoe, M.~M.~Nojiri and G.~Polesello,
  ``A new SUSY mass reconstruction method at the CERN LHC,''
  Phys.\ Rev.\  D {\bf 71}, 035008 (2005)
  [arXiv:hep-ph/0410160].

\bibitem{Gjelsten:2004ki}
  B.~K.~Gjelsten, D.~J.~Miller and P.~Osland,
  ``Measurement of SUSY masses via cascade decays for SPS 1a,''
  JHEP {\bf 0412}, 003 (2004)
  [arXiv:hep-ph/0410303].

\bibitem{Gjelsten:2005aw}
  B.~K.~Gjelsten, D.~J.~Miller and P.~Osland,
  ``Measurement of the gluino mass via cascade decays for SPS 1a,''
  JHEP {\bf 0506}, 015 (2005)
  [arXiv:hep-ph/0501033].

\bibitem{Miller:2005zp}
  D.~J.~Miller, P.~Osland and A.~R.~Raklev,
  ``Invariant mass distributions in cascade decays,''
  JHEP {\bf 0603}, 034 (2006)
  [arXiv:hep-ph/0510356].

\bibitem{Lester:2006yw}
  C.~G.~Lester,
  ``Constrained invariant mass distributions in cascade decays: The shape of
  the 'm(qll)-threshold' and similar distributions,''
  Phys.\ Lett.\  B {\bf 655}, 39 (2007)
  [arXiv:hep-ph/0603171].

\bibitem{Gjelsten:2006tg}
  B.~K.~Gjelsten, D.~J.~Miller, P.~Osland and A.~R.~Raklev,
  ``Mass determination in cascade decays using shape formulas,''
  AIP Conf.\ Proc.\  {\bf 903}, 257 (2007)
  [arXiv:hep-ph/0611259].

\bibitem{Matsumoto:2006ws}
  S.~Matsumoto, M.~M.~Nojiri and D.~Nomura,
  ``Hunting for the top partner in the littlest Higgs model with T-parity at
  the LHC,''
  Phys.\ Rev.\  D {\bf 75}, 055006 (2007)
  [arXiv:hep-ph/0612249].

\bibitem{Cheng:2007xv}
  H.~C.~Cheng, J.~F.~Gunion, Z.~Han, G.~Marandella and B.~McElrath,
  ``Mass Determination in SUSY-like Events with Missing Energy,''
  JHEP {\bf 0712}, 076 (2007)
  [arXiv:0707.0030 [hep-ph]].

\bibitem{Lester:2007fq}
  C.~Lester and A.~Barr,
  ``MTGEN : Mass scale measurements in pair-production at colliders,''
  JHEP {\bf 0712}, 102 (2007)
  [arXiv:0708.1028 [hep-ph]].

\bibitem{Cho:2007qv}
  W.~S.~Cho, K.~Choi, Y.~G.~Kim and C.~B.~Park,
  ``Gluino Stransverse Mass,''
  Phys.\ Rev.\ Lett.\  {\bf 100}, 171801 (2008)
  [arXiv:0709.0288 [hep-ph]].

\bibitem{Gripaios:2007is}
  B.~Gripaios,
  ``Transverse Observables and Mass Determination at Hadron Colliders,''
  JHEP {\bf 0802}, 053 (2008)
  [arXiv:0709.2740 [hep-ph]].

\bibitem{Barr:2007hy}
  A.~J.~Barr, B.~Gripaios and C.~G.~Lester,
   ``Weighing Wimps with Kinks at Colliders: Invisible Particle Mass
  Measurements from Endpoints,''
  JHEP {\bf 0802}, 014 (2008)
  [arXiv:0711.4008 [hep-ph]].

\bibitem{Cho:2007dh}
  W.~S.~Cho, K.~Choi, Y.~G.~Kim and C.~B.~Park,
   ``Measuring superparticle masses at hadron collider using the transverse mass
  kink,''
  JHEP {\bf 0802}, 035 (2008)
  [arXiv:0711.4526 [hep-ph]].

\bibitem{Ross:2007rm}
  G.~G.~Ross and M.~Serna,
  ``Mass Determination of New States at Hadron Colliders,''
  Phys.\ Lett.\  B {\bf 665}, 212 (2008)
  [arXiv:0712.0943 [hep-ph]].

\bibitem{Nojiri:2007pq}
  M.~M.~Nojiri, G.~Polesello and D.~R.~Tovey,
   ``A hybrid method for determining SUSY particle masses at the LHC with fully
  identified cascade decays,''
  JHEP {\bf 0805}, 014 (2008)
  [arXiv:0712.2718 [hep-ph]].

\bibitem{Huang:2008ae}
  P.~Huang, N.~Kersting and H.~H.~Yang,
  ``Hidden Thresholds: A Technique for Reconstructing New Physics Masses at
  Hadron Colliders,''
  arXiv:0802.0022 [hep-ph].

\bibitem{Nojiri:2008hy}
  M.~M.~Nojiri, Y.~Shimizu, S.~Okada and K.~Kawagoe,
  ``Inclusive transverse mass analysis for squark and gluino mass
  determination,''
  JHEP {\bf 0806}, 035 (2008)
  [arXiv:0802.2412 [hep-ph]].

\bibitem{Tovey:2008ui}
  D.~R.~Tovey,
  ``On measuring the masses of pair-produced semi-invisibly decaying particles
  at hadron colliders,''
  JHEP {\bf 0804}, 034 (2008)
  [arXiv:0802.2879 [hep-ph]].

\bibitem{Nojiri:2008ir}
  M.~M.~Nojiri and M.~Takeuchi,
  ``Study of the top reconstruction in top-partner events at the LHC,''
  arXiv:0802.4142 [hep-ph].

\bibitem{Cheng:2008mg}
  H.~C.~Cheng, D.~Engelhardt, J.~F.~Gunion, Z.~Han and B.~McElrath,
  ``Accurate Mass Determinations in Decay Chains with Missing Energy,''
  Phys.\ Rev.\ Lett.\  {\bf 100}, 252001 (2008)
  [arXiv:0802.4290 [hep-ph]].

\bibitem{Cho:2008cu}
  W.~S.~Cho, K.~Choi, Y.~G.~Kim and C.~B.~Park,
  ``Measuring the top quark mass with $m_{T2}$ at the LHC,''
  Phys.\ Rev.\  D {\bf 78}, 034019 (2008)
  [arXiv:0804.2185 [hep-ph]].

\bibitem{Serna:2008zk}
  M.~Serna,
  ``A short comparison between $m_{T2}$ and $m_{CT}$,''
  JHEP {\bf 0806}, 004 (2008)
  [arXiv:0804.3344 [hep-ph]].

\bibitem{Bisset:2008hm}
  M.~Bisset, R.~Lu and N.~Kersting,
  ``Improving SUSY Spectrum Determinations at the LHC with Wedgebox and Hidden
  Threshold Techniques,''
  arXiv:0806.2492 [hep-ph].

\bibitem{Barr:2008ba}
  A.~J.~Barr, G.~G.~Ross and M.~Serna,
  ``The Precision Determination of Invisible-Particle Masses at the LHC,''
  arXiv:0806.3224 [hep-ph].

\bibitem{Kersting:2008qn}
  N.~Kersting,
  ``On Measuring Split-SUSY Gaugino Masses at the LHC,''
  arXiv:0806.4238 [hep-ph].

\bibitem{Nojiri:2008vq}
  M.~M.~Nojiri, K.~Sakurai, Y.~Shimizu and M.~Takeuchi,
  ``Handling jets + missing $E_T$ channel using inclusive mT2,''
  arXiv:0808.1094 [hep-ph].

\bibitem{Alwall:2008va}
  J.~Alwall, M.~P.~Le, M.~Lisanti and J.~G.~Wacker,
  ``Model-Independent Jets plus Missing Energy Searches,''
  arXiv:0809.3264 [hep-ph].

\bibitem{Cho:2008tj}
  W.~S.~Cho, K.~Choi, Y.~G.~Kim and C.~B.~Park,
  ``$M_{T2}$-assisted on-shell reconstruction of missing momenta and its
  application to spin measurement at the LHC,''
  arXiv:0810.4853 [hep-ph].

\bibitem{Cheng:2008hk}
  H.~C.~Cheng and Z.~Han,
  ``Minimal Kinematic Constraints and MT2,''
  arXiv:0810.5178 [hep-ph].

\bibitem{Burns:2008va}
  M.~Burns, K.~Kong, K.~T.~Matchev and M.~Park,
  ``Using Subsystem MT2 for Complete Mass Determinations in Decay Chains with
  Missing Energy at Hadron Colliders,''
  arXiv:0810.5576 [hep-ph].

\bibitem{Barr:2008hv}
  A.~J.~Barr, A.~Pinder and M.~Serna,
  ``Precision Determination of Invisible-Particle Masses at the CERN LHC: II,''
  arXiv:0811.2138 [hep-ph].

\bibitem{Konar:2008ei}
  P.~Konar, K.~Kong and K.~T.~Matchev,
  ``$\sqrt{s}_{\rm min}$: a global inclusive variable for determining the mass scale of
  new physics in events with missing energy at hadron colliders,''
  arXiv:0812.1042 [hep-ph].

\bibitem{Martin:2009iq}
  A.~D.~Martin, W.~J.~Stirling, R.~S.~Thorne and G.~Watt,
  ``Parton distributions for the LHC,''
  arXiv:0901.0002 [hep-ph].

\bibitem{Corcella:2000bw}
  G.~Corcella {\it et al.},
  ``HERWIG 6.5: an event generator for Hadron Emission Reactions With
  Interfering Gluons (including supersymmetric processes),''
  JHEP {\bf 0101} (2001) 010
  [arXiv:hep-ph/0011363].

\bibitem{Corcella:2002jc}
  G.~Corcella {\it et al.},
  ``HERWIG 6.5 release note,''
  arXiv:hep-ph/0210213.

\bibitem{GETJET} F.~Paige, private communication; F.~Paige and S.~Protopopescu,
``ISAJET 5.02: a Monte Carlo event generator for $pp$ and $\bar p p$ interactions,''
in {\it Supercollider Physics}, ed.\ D.~Soper (World Scientific, Singapore, 1986), p.~41.

\bibitem{Mangano:2006rw}
  M.~L.~Mangano, M.~Moretti, F.~Piccinini and M.~Treccani,
  ``Matching matrix elements and shower evolution for top-quark production in
  hadronic collisions,''
  JHEP {\bf 0701} (2007) 013
  [arXiv:hep-ph/0611129].

\bibitem{Alwall:2008qv}
  J.~Alwall, S.~de Visscher and F.~Maltoni,
  ``QCD radiation in the production of heavy colored particles at the LHC,'''
  JHEP {\bf 0902} (2009) 017
  [arXiv:0810.5350 [hep-ph]].

\bibitem{Hoeche:2009rj}
  S.~Hoeche, F.~Krauss, S.~Schumann and F.~Siegert,
  ``QCD matrix elements and truncated showers,''
  JHEP {\bf 0905} (2009) 053
  [arXiv:0903.1219 [hep-ph]].

\bibitem{Frixione:2002ik}
  S.~Frixione and B.~R.~Webber,
  ``Matching NLO QCD computations and parton shower simulations,''
  JHEP {\bf 0206} (2002) 029
  [arXiv:hep-ph/0204244].

\bibitem{Frixione:2003ei}
  S.~Frixione, P.~Nason and B.~R.~Webber,
  ``Matching NLO QCD and parton showers in heavy flavour production,''
  JHEP {\bf 0308} (2003) 007
  [arXiv:hep-ph/0305252].

\end{thebibliography}
\end{document}